\documentclass[pdflatex,sn-nature]{sn-jnl}% Style for submissions to Nature Portfolio journals % ,referee

\usepackage[separate-uncertainty=true, multi-part-units=single, exponent-product=\cdot]{siunitx}
\newcommand{\dd}{\text{d}}
\usepackage{placeins}

\usepackage{graphicx}%
\usepackage{multirow}%
\usepackage{amsmath,amssymb,amsfonts}%
\usepackage{amsthm}%
\usepackage{mathrsfs}%
\usepackage[title]{appendix}%
\usepackage{xcolor}%
\usepackage{textcomp}%
\usepackage{manyfoot}%
\usepackage{booktabs}%
\usepackage{algorithm}%
\usepackage{algorithmicx}%
\usepackage{algpseudocode}%
\usepackage{listings}%
\usepackage[displaymath, mathlines]{lineno}

\newgeometry{bindingoffset=11mm, marginparsep=0mm, marginparwidth=0mm}

\begin{document}

\title{Limited impact of Greenland meltwater on abruptness and reversibility of future Atlantic overturning changes}

\author*[1,2]{\fnm{Oliver} \sur{Mehling}}\email{o.m.mehling@uu.nl}

\author[1,3]{\fnm{Katinka} \sur{Bellomo}}

\author[4]{\fnm{Federico} \sur{Fabiano}}

\author[5]{\fnm{Marion} \sur{Devilliers}}

\author[6]{\fnm{Michele} \sur{Petrini}}

\author[4]{\fnm{Susanna} \sur{Corti}}

\author[1,7]{\fnm{Jost} \spfx{von} \sur{Hardenberg}}

\affil[1]{Politecnico di Torino, Department of Environment, Land and Infrastructure Engineering, Torino, Italy}
\affil[2]{Utrecht University, Institute for Marine and Atmospheric Research, Utrecht, The Netherlands}
\affil[3]{University of Padova, Department of Geosciences, Padua, Italy}
\affil[4]{Consiglio Nazionale delle Ricerche, Institute of Atmospheric Sciences and Climate, Bologna, Italy}
\affil[5]{Danish Meteorological Institute, Research and Development, Copenhagen, Denmark}
\affil[6]{NORCE Norwegian Research Centre, Bjerknes Centre for Climate Research, Bergen, Norway}
\affil[7]{Consiglio Nazionale delle Ricerche, Institute of Atmospheric Sciences and Climate, Turin, Italy}

\abstract{All climate models project that the Atlantic Meridional Overturning Circulation (AMOC) will weaken in the 21\textsuperscript{st} century, but most models neglect increasing runoff from the Greenland ice sheet. Greenland meltwater is expected to exacerbate AMOC weakening, and omitting it increases the uncertainty in assessing the possibility of a future abrupt collapse or tipping of the AMOC. Here, we test the abruptness and reversibility of AMOC changes under strong future global warming in a state-of-the-art climate model with and without physically plausible Greenland meltwater forcing. While Greenland meltwater significantly exacerbates future AMOC weakening, modeled long-term AMOC changes are neither abrupt nor irreversible. While accounting for Greenland meltwater will increase the accuracy of climate projections, our results suggest that the importance of Greenland meltwater for assessing the risk of future AMOC tipping may be smaller than previously thought.}

\maketitle

%\graphicspath{{Figures}}

\linespread{1.5}\selectfont
%\linenumbers

\section{Introduction}\label{sec:intro}
The Atlantic Meridional Overturning Circulation (AMOC) has a prominent role in shaping the Earth's mean climate and climate change \cite{Buckley2016,Liu2020,Bellomo2021,Hu2025}. For instance, recent studies have demonstrated the influence of an AMOC weakening on the position of the inter-tropical convergence zone \cite{Cerato2025}, European temperature extremes \cite{Meccia2024,Meccia2025,VanWesten2025a}, Arctic amplification \cite{Lee2024}, Pacific trade winds \cite{Orihuela-Pinto2022}, and global monsoon systems \cite{Ben-Yami2024b}.

Because of these near-global impacts, the possibility that global warming could induce an abrupt collapse or \textit{tipping} of the AMOC, defined here as ``a critical threshold beyond which a system reorganizes, often abruptly and/or irreversibly'' \cite{Chen2021}, has been debated for several decades \cite{Broecker1985,Weijer2019,ArmstrongMcKay2022}. Early warning indicators suggest that the AMOC is currently undergoing destabilization compatible with approaching a tipping point \cite{Boers2021,Ditlevsen2023,VanWesten2024}, although recent work has also identified new mechanisms that are expected to stabilize the AMOC \cite{Baker2025,Bonan2025}. In addition, the ability of the AMOC to weaken abruptly and irreversibly with respect to (idealized) freshwater forcing has now been demonstrated across the hierarchy of ocean and climate models \cite{Stommel1961,Rahmstorf2005,VanWesten2024,VanWesten2025}.

Nevertheless, while there is consensus among climate models that the AMOC will weaken until 2100 \cite{Weijer2020}, the IPCC Sixth Assessment Report (AR6) found that none of the CMIP6 models show an abrupt AMOC collapse during the 21\textsuperscript{st} century \cite{Fox-Kemper2021}. In contrast to earlier assessments, however, only \textit{medium confidence} was assigned to this finding, implying that a possible collapse of the AMOC is still under debate \cite{Fox-Kemper2021}. Besides model biases that could influence AMOC stability \cite{Liu2017,VanWesten2024b}, one main reason why current-generation models may not be able to simulate an AMOC collapse is that they neglect meltwater influx from the Greenland Ice Sheet \cite{Fox-Kemper2021}, which is expected to accelerate with global warming \cite{Trusel2018}.

Previous studies have shown that including realistic Greenland ice sheet meltwater in future emission scenarios can exacerbate future AMOC weakening \cite{Fichefet2003,Ridley2005,Mikolajewicz2007,Gierz2015,Lenaerts2015,Bakker2016,Golledge2019,Ackermann2020}, but the impact of Greenland meltwater on abruptness and irreversibility has not yet been investigated, even though meltwater is often hypothesized as a mechanism that would induce or accelerate a future AMOC collapse \cite{Bakker2016,Caesar2018,Fox-Kemper2021,Drijfhout2025}. Here, we assess the long-term AMOC response until 2300 under strong greenhouse gas emissions and Greenland meltwater forcing in model experiments carried out with current-generation (CMIP6) climate model EC-Earth3 \cite{Doscher2022}. This model features one of the higher grid resolutions in CMIP6 and is one of the few CMIP6 models which simulates both a realistic present-day AMOC strength and a negative value of freshwater import into the South Atlantic ($F_{ovS}$; Methods), which is consistent with observations and implies that the model is not expected to be biased towards a too stable AMOC \cite{VanWesten2024b}. Despite a significant impact of Greenland meltwater on AMOC weakening especially after 2100, we do not find an abrupt or irreversible AMOC collapse, suggesting that the omission of Greenland mass loss in climate models might impact the assessment of a potential future AMOC tipping less than previously thought.

\section{Results}\label{sec2}

\subsection{Meltwater-induced AMOC weakening}

\begin{figure}[htbp]
    \centering
    \includegraphics[width=0.8\textwidth]{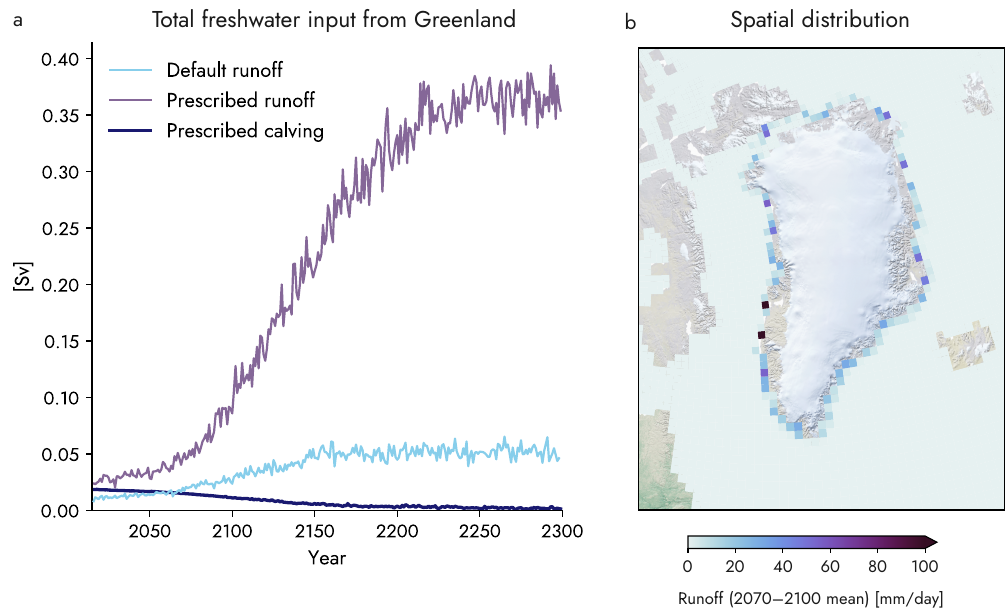}
    \caption{\textbf{Greenland meltwater forcing.} (a) Annual mean runoff and calving used as input for the ``Meltwater'' simulations and runoff in the ``Reference'' simulations for comparison, (b) Example spatial distribution of runoff (averaged over 2070--2100) with non-uniform input at coastal grid points.}
    \label{fig:fw-forcing}
\end{figure}

To reliably isolate and quantify the meltwater signal, we perform two four-member initial condition ensembles with and without the Greenland meltwater forcing (``Meltwater'' and ``Reference'', respectively). We prescribe Greenland runoff and calving from a state-of-the-art (CMIP6) coupled climate--ice sheet model simulation \cite{Muntjewerf2020} under a high-end greenhouse gas emission scenario (SSP5-8.5) until 2300 (Methods). Meltwater derived from the coupled climate--ice sheet model reaches \SI{0.09}{Sv} by 2100 and more than \SI{0.3}{Sv} by 2300 (Fig. \ref{fig:fw-forcing}a), larger values than those considered in previous studies with parametrized meltwater (see Methods). In contrast to most previous studies, the meltwater is also routed in a physically consistent way, non-uniformly in space and time, to coastal grid points around Greenland (Fig. \ref{fig:fw-forcing}b).

\begin{figure}[htbp]
    \centering
    \includegraphics[width=\textwidth]{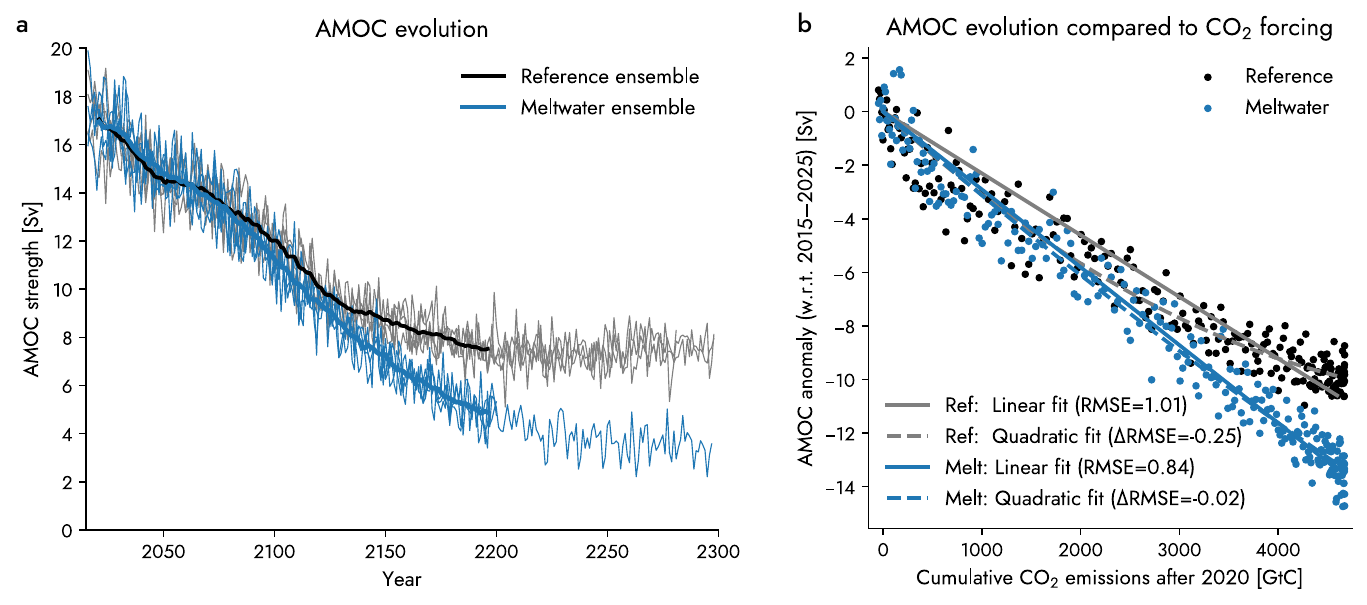}
    \caption{\textbf{AMOC projections until 2300 with and without Greenland meltwater.} (a) Annual mean AMOC time series for individual ensemble members (thin lines) and 10-year running mean of ensemble means until 2200 (thick lines; as the full 4$\times$2 member ensemble covers the period 2016--2200), (b) AMOC anomalies in one pair of ensemble members against cumulative CO$_2$ emissions since 2015.}
    \label{fig:amoc-ts}
\end{figure}

We find that Greenland meltwater consistently exacerbates the weakening of the AMOC at \ang{26.5}N in response to global warming (Fig. \ref{fig:amoc-ts}a). However, until 2100, the size of the effect is projected to be relatively small. For the reference ensemble without Greenland meltwater, the AMOC weakening under SSP5-8.5 is approximately linear during the 21\textsuperscript{st} century and ranges between \SI{3.8}{} and \SI{7.3}{Sv} per century for different ensemble members, compared to \SI{4.5}{} to \SI{8.4}{Sv} per century in the meltwater ensemble. This enhanced weakening of, on average, \SI{0.9}{Sv} per century is consistent for all pairs of ensemble members (Supplementary Fig. \ref{fig:amoc-trends-21st}a) and statistically significant ($p = 0.02$ using a paired Student's $t$-test), although on the same order as the standard deviation of trends due to low-frequency variability (\SI{1.5}{Sv} per century in the reference ensemble).

We quantify when this significantly stronger AMOC weakening trend emerges in the meltwater ensemble compared to the reference ensemble using bootstrap resampling of the interannual to decadal AMOC variability (Methods). For the linear AMOC trend, a statistically significant difference emerges in 2092 (2086 to 2098, 66\% confidence interval [CI]; Fig. \ref{fig:amoc-ts}b). Similarly, an emergence in 2100 (2091 to 2112, 66\% CI) is found when the 15-year running mean of the AMOC index is used instead of the AMOC trend. This is at most a few decades after first differences between forcing scenarios emerge, given that the AMOC in CMIP6 models weakens almost independently of the scenario ''until about 2060'' \cite{Fox-Kemper2021}.

Beyond the 21\textsuperscript{st} century, the meltwater-induced AMOC weakening becomes stronger, in line with the strong increase in Greenland melting after 2100 (Fig. \ref{fig:fw-forcing}) due to the sustained high CO$_2$ emissions in the SSP5-8.5 scenario. By the end of the 22\textsuperscript{nd} century (2180-2200), an additional AMOC weakening of \SI{2.5(2)}{Sv} is attributable to Greenland meltwater. We extend one member per ensemble until the end of the 23\textsuperscript{rd} century, by which the AMOC stabilizes at \SI{7.5}{Sv} in the reference simulation but weakens further to about \SI{3.5}{Sv} with added Greenland meltwater. 
In summary, Greenland meltwater induces an additional AMOC weakening on the order of 10--20\% at the end of the 21\textsuperscript{st} century and up to 40\% at the end of the 23\textsuperscript{rd} century, compared to the AMOC weakening induced by CO$_2$ and consequent atmosphere--ocean feedbacks which are accounted for in current climate models.

Even after experiencing strong meltwater input, a basin-wide Atlantic overturning cell remains in all simulations, although it is shallower and weaker in the meltwater simulation (Fig. \ref{fig:amoc-streamfunctions-23rd}). This differs from entirely collapsed or reversed Atlantic overturning states previously identified in some climate models \cite{Manabe1999,Gregory2003,VanWesten2023}. Nevertheless, the AMOC in the meltwater simulations is on the threshold of being classified as ``collapsed'' using typical threshold definitions (e.g., 80\% weakening compared to the pre-industrial AMOC strength \cite{Collins2019}). AMOC characteristics such as northward heat transport due to overturning into the South Atlantic (Supplementary Fig. \ref{fig:characteristics-amoc-2300}a) or shared outcropping isopycnals between the Northern hemisphere and the Southern Ocean \cite{Nikurashin2012} (Supplementary Fig. \ref{fig:characteristics-amoc-2300}b,c) remain even at the end of the 23\textsuperscript{rd} century, but they are strongly reduced with meltwater input. It should also be noted that the Atlantic ocean heat transport due to overturning is indeed close to zero at some tropical latitudes in the meltwater simulation (not shown).

More importantly, we do not find signs of an \textit{abrupt} AMOC change following the classical definition of a nonlinear response exceeding the rate of external forcing \cite{Rahmstorf2001,Alley2003}. As shown in Fig. \ref{fig:amoc-ts}b, the AMOC strength in the meltwater ensemble scales remarkably linearly with cumulative CO$_2$ emissions until emissions reach zero in 2250. This scaling is sub-linear in the reference ensemble due to the AMOC stabilization in the 23\textsuperscript{rd} century, in agreement with previous modeling studies which did not account for GrIS melt \cite{Steinacher2016}. To summarize, while it is ambiguous whether the AMOC can be classified as ``collapsed'', the weakening is not abrupt.

\subsection{Shift of source regions shapes response to meltwater}
\begin{figure}[htbp]
    \centering
    \includegraphics[width=0.8\textwidth]{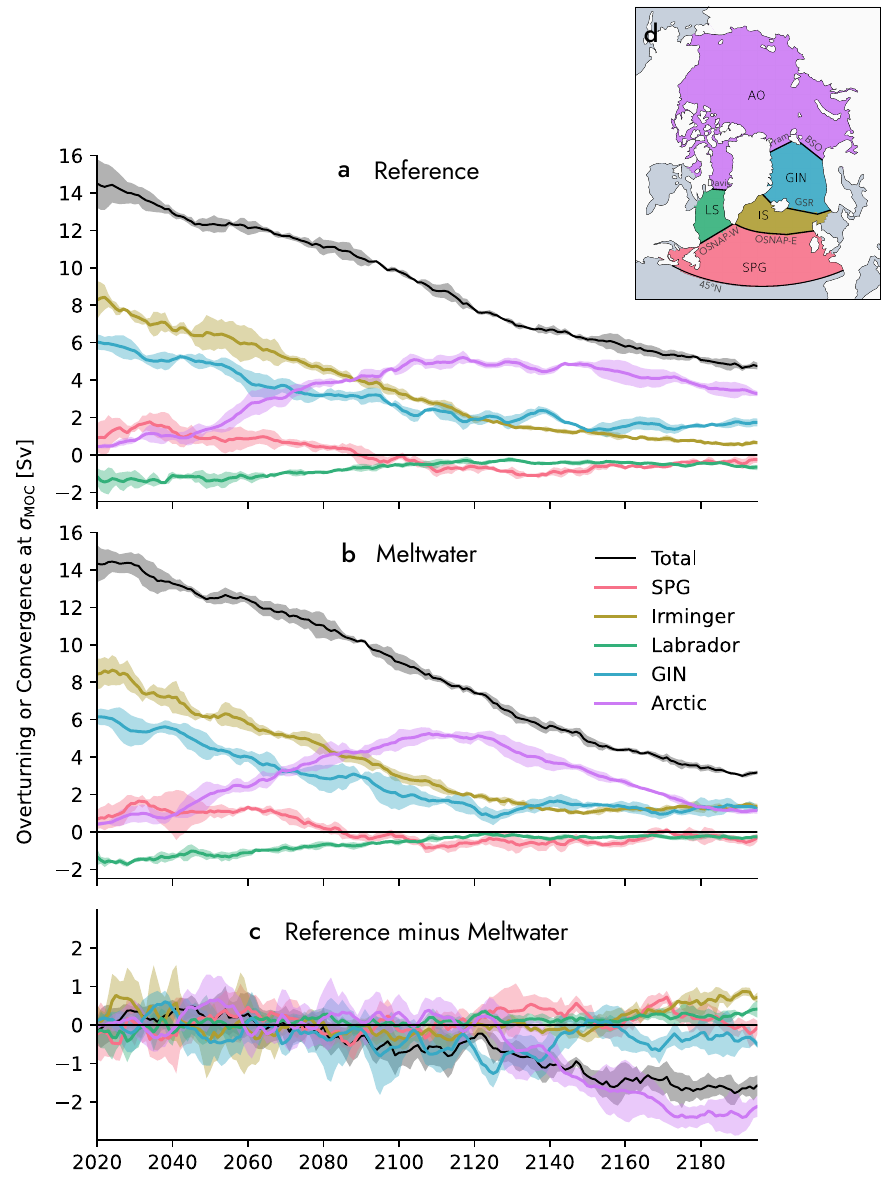}
    \caption{\textbf{Change in AMOC source regions.} Maximum overturning at \ang{45}N (black line) and convergence at $\sigma_\text{MOC}$ by region: (a) Reference ensemble, (b) Meltwater ensemble, (c) Difference between the two ensembles. All lines are for 2015--2200 and are smoothed with a ten-year running mean; shadings indicate the ensemble standard deviation. The inset (d) shows the definition of the source regions and the gateways bounding them.}
    \label{fig:convergence-sigma-moc}
\end{figure}

The significant additional AMOC weakening after around 2100 calls for a better understanding of associated mechanisms. To this end, we decompose the Atlantic overturning at \ang{45}N and its meltwater-induced weakening into the contributions of different source regions north of \ang{45}N (defined in Fig. \ref{fig:convergence-sigma-moc}d) based on the density-space overturning at the northern and southern gateways bounding these regions (Methods).

Both ensembles agree that the main source regions of the present-day AMOC in EC-Earth3, the Irminger and Nordic Seas, will decrease in importance through the 21\textsuperscript{st} and 22\textsuperscript{nd} centuries. In turn, the AMOC is characterized by a strong northward shift in source regions in response to global warming. The Arctic Ocean, whose present-day direct contribution to the AMOC at \ang{45}N is negligible, becomes the most important AMOC source region by 2100 (Fig. \ref{fig:convergence-sigma-moc}a--b). This increased role of the Arctic Ocean is partly because the overturning strength across the Arctic gateways intensifies, but also because the overturning peak in density space moves more rapidly to lighter densities in the Arctic Ocean than at subpolar latitudes (Fig. \ref{fig:moc-convergence-snapshots}), potentially due to Arctic Ocean amplification \cite{Shu2022}. Consequently, the density of maximum overturning at the Arctic gateways coincides with that at \ang{45}N $\sigma_\text{MOC}$ during much of the 22\textsuperscript{nd} century, whereas it is located at the larger densities of the overflow waters during the present-day, contributing to the overturning further south only indirectly (e.g., via mixing at the Greenland--Scotland Ridge).

\begin{figure}[p]
    \centering
    \includegraphics[width=0.8\textwidth]{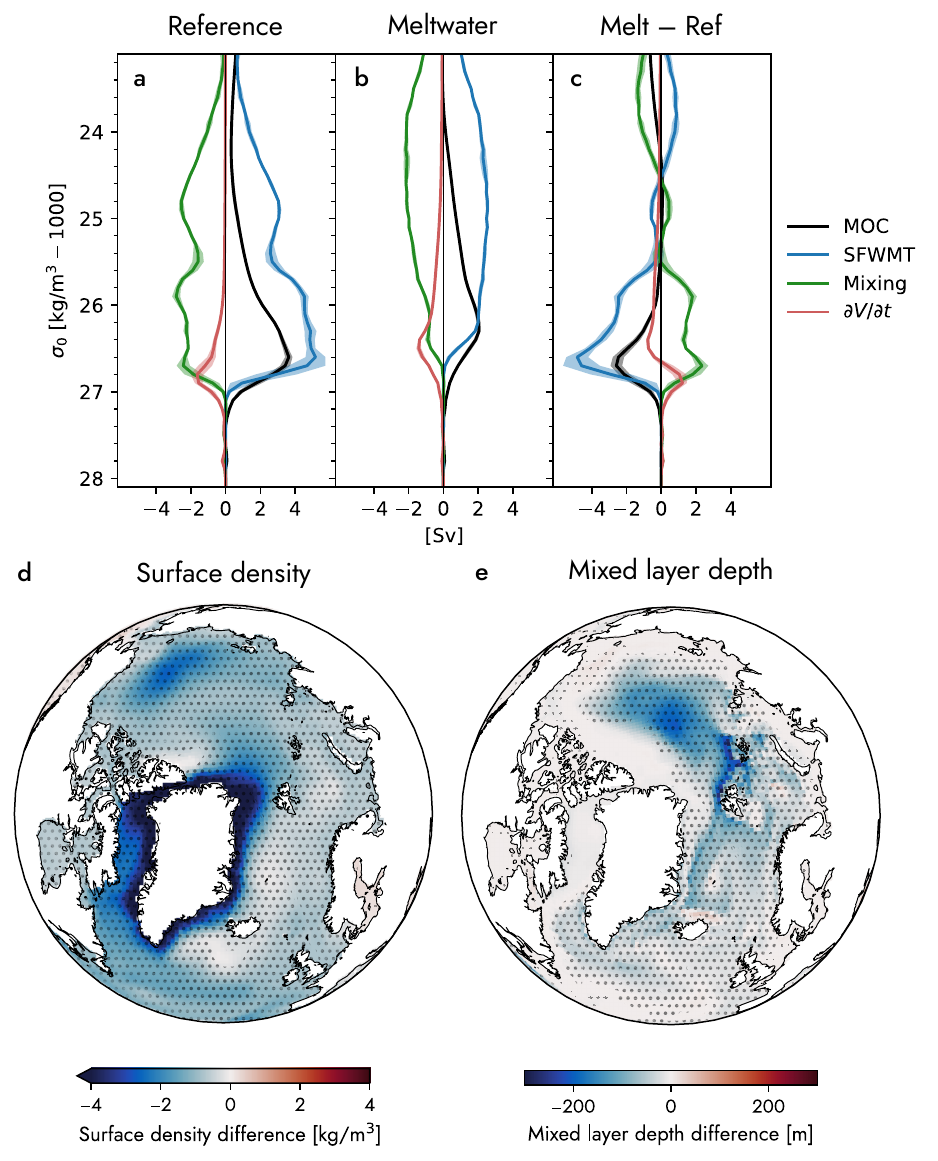}
    \caption{\textbf{Effect of Greenland meltwater on an ice-free Arctic Ocean.} a--c: Volume and buoyancy budget (Methods; \cite{Buckley2023}) for the Arctic Ocean at the end of the 22\textsuperscript{nd} century for (a) the reference simulation, (b) with Greenland meltwater, (c) difference between b and a. The overturning across the Arctic gateways (MOC) is decomposed into surface-forced water mass transformation (SFWMT), volume changes ($\partial V/\partial t$) and mixing, which is calculated as a residual. d--e: Ensemble mean differences (meltwater minus reference) at the end of the 22\textsuperscript{nd} century: (d) annual mean surface density, (e) winter (March) mixed layer depth. Areas with significant differences ($p<0.05$) using a $t$-test are stippled.}
    \label{fig:arctic-ocean-end22nd}
\end{figure}

In line with the northward shift of AMOC source regions, the Arctic Ocean contributes most to the meltwater-induced additional AMOC weakening in the 22\textsuperscript{nd} century (Fig. \ref{fig:convergence-sigma-moc}c). This meltwater-driven weakening can be understood in terms of a volume and buoyancy budget for the Arctic Ocean (Fig \ref{fig:arctic-ocean-end22nd}a-c; Methods). The overturning across the Arctic gateways weakens mostly due to a decrease in surface-forced water mass transformation (SFWMT), while the magnitude of interior mixing also decreases with meltwater input. Decreased SFWMT is linked to a decrease in both surface density (Fig. \ref{fig:arctic-ocean-end22nd}d) and surface buoyancy flux. While the former is a direct consequence of the meltwater input freshening the surface layers, the latter (as well as the smaller mixing component) is linked to a decrease in mixed layer depth (Fig. \ref{fig:arctic-ocean-end22nd}e). This is again a consequence of the lower surface density, which increases stratification. To summarize, Greenland meltwater influences the northward-shifted AMOC source regions both directly (via decreased surface density) and indirectly (via increased stratification that leads to less mixing and less heat loss).

To gauge the robustness of meltwater impacts, we calculate the Arctic Ocean contribution to the overturning at $\sigma_{\text{MOC}}$ in eight other CMIP6 models which provide SSP5-8.5 simulations until 2300 (Fig. \ref{fig:convergence-arctic-cmip6}). In all models, the Arctic contribution to Atlantic overturning increases during the 21\textsuperscript{st} century, in agreement with EC-Earth3. However, the timing of the maximum contribution is model-dependent. In most models, the Arctic contribution decreases to below \SI{2}{Sv} by the mid-22\textsuperscript{nd} century, while EC-Earth3 is the only model from this ensemble with a sustained Arctic contribution until 2300. This implies that EC-Earth3 probably provides an upper-end estimate of the meltwater-induced AMOC weakening via the Arctic Ocean after 2150, although other regions might play a significant role in modulating meltwater-induced AMOC weakening in other models.

\subsection{Reversibility of meltwater-induced AMOC changes}

To test the impact of the added Greenland meltwater on the reversibility of AMOC changes, we branch off idealized reversibility experiments in which the meltwater and/or greenhouse gas forcings are reversed. We conduct two sets of experiments starting from the year 2250, in which emissions reach zero in the extended SSP5-8.5 scenario. In the first set of experiments, the meltwater forcing is switched off in 2250 and the CO$_2$ concentration is kept constant after 2300. This enables assessing AMOC reversibility with respect to meltwater under sustained strong global warming. In the second set of experiments, CO$_2$ concentrations are ramped down at 1\% per year for 170 years until they reach 2015 levels and held constant thereafter. This ramp-down experiment is initialized from both the meltwater and reference experiments to isolate differences in the reversibility behavior due to the previous meltwater forcing. Because no CO$_2$ ramp-down is available with the ice sheet model, we also switch off the additional meltwater forcing after 2250 for these CO$_2$ reversal experiments.

\begin{figure}[htbp]
    \centering
    \includegraphics[width=0.9\textwidth]{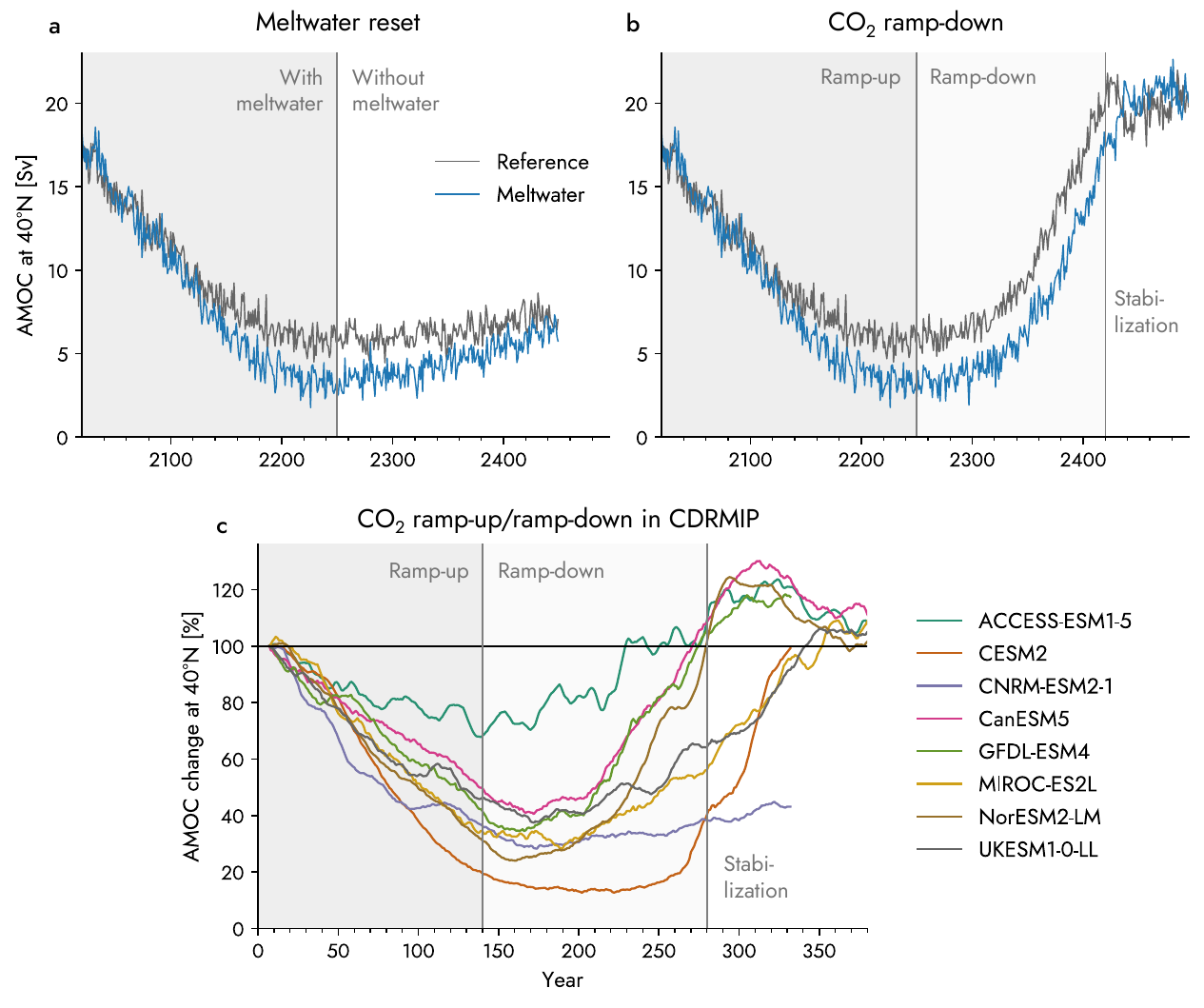}
    \caption{\textbf{AMOC reversibility in EC-Earth3 and CMIP6 models.} Annual mean AMOC maximum at \ang{40}{N} for (a) the ``meltwater reset'' experiments, (b) the ``CO$_2$ ramp-down'' experiments (see the text for experiment setups). The dark grey-shaded time spans in both subplots are identical to the standard ``Reference'' and ``Meltwater'' simulations, with the meltwater forcing stopped at 2250. (c) AMOC change in different CMIP6 models in the CDRMIP reversibility experiments \cite{Keller2018}. Starting from pre-industrial conditions, these experiment consist of a CO$_2$ ramp-up at 1\% per year for 140 years, followed by a ramp-down at 1\% per year for another 140 years and a stabilization period at pre-industrial values for at least 50 years. AMOC changes are indicated as a percentage of the initial (year 0--15) AMOC strength and smoothed using a 15-year running mean.}
    \label{fig:reversibility}
\end{figure}

Resetting the meltwater forcing under late-23\textsuperscript{rd} century conditions leads to a gradual recovery of the AMOC (Fig. \ref{fig:reversibility}a). We continue the simulation for 200 years, after which the AMOC anomaly at \ang{40}N has decreased from \SI{-2.7}{Sv} to \SI{-0.7}{Sv} compared to the reference simulation, which is extended using the same constant CO$_2$ concentrations after 2300. The larger recovery rate in the meltwater simulation suggests that both simulations would eventually converge to the same equilibrium. Therefore, while the timescale of recovery is (multi-)centennial, it appears that Greenland meltwater forcing, despite its large magnitude, does not induce an irreversible shift to a different equilibrium state of the AMOC under very high CO$_2$ concentrations.

Similarly, the CO$_2$ ramp-down experiments show reversibility of both the greenhouse gas- and meltwater-induced AMOC changes (Fig. \ref{fig:reversibility}b). During the ramp-down, the AMOC is about 2--3 Sv stronger without prior meltwater forcing, with the main regional driver of differences shifting from the Arctic to the Nordic Seas later during the ramp-down (Supplementary Fig. \ref{fig:convergence-reversibility}). This corresponds to a difference in recovery time to the same AMOC strength of, on average, 25 years. Both simulations eventually approach a similar stabilized AMOC strength of around \SI{20}{Sv} after CO$_2$ concentrations have reached 2015 levels. This is stronger than the initial AMOC strength in 2015, in agreement with previous CO$_2$ ramp-down experiments without meltwater \cite{Jackson2014}. In EC-Earth3, this could be linked to a permanently winter ice-free Labrador Sea, whereas the Labrador Sea is intermittently ice-covered in the pre-industrial \cite{Mehling2024a} and historical \cite{Doscher2022} simulations. This points toward a potential irreversibility of winter sea ice changes \cite{Yu2025}, which will be investigated elsewhere.

The reversibility to a ramp-down of CO$_2$ concentrations is not unique to EC-Earth3, but can be found across the ensemble of CMIP6 models in the \textit{1pctCO2-cdr} simulations of the CDRMIP project \cite{Keller2018}. These simulations followed a similar CO$_2$ ramp-down protocol at 1\% per year from four times pre-industrial (Methods), albeit without explicitly considering Greenland meltwater. In the CDRMIP simulations, the AMOC fully recovers in 7 out of 8 CMIP6 models \cite{Schwinger2022} (Fig. \ref{fig:reversibility}c). Our simulations suggest that this reversibility might not be sensitive to prior Greenland meltwater input, even when its magnitude is very large.

\section{Discussion} \label{sec:discussion}
In this study, we used an ensemble of future projections with a state-of-the-art climate model to probe the response of the future AMOC to high-end but physically plausible Greenland meltwater input. In contrast to a non-significant effect during the historical period \cite{Devilliers2024}, we found a small but significant additional AMOC weakening associated with Greenland meltwater of about \SI{1}{Sv} until 2100 (about 10\% of the CO$_2$-induced weakening) and up to \SI{4}{Sv} until 2300 (nearly 40\% of the CO$_2$-induced weakening) under very strong forcing. These results are in good agreement with previous studies using parametrized Greenland meltwater input \cite{Lenaerts2015,Bakker2016}. However, we found that the combined CO$_2$- and meltwater-induced AMOC weakening in our model was neither abrupt (with respect to the external forcing) nor irreversible, characteristics often associated with climate tipping points \cite{Chen2021,ArmstrongMcKay2022}. Instead, with Greenland meltwater, the AMOC weakening scaled remarkably linearly with cumulative CO$_2$ emissions. This linearity, if corroborated by other climate models, might also yield additional insights into the physics of future AMOC weakening, such as why the AMOC strength diverges relatively late in different emission scenarios \cite{Weijer2020}.

As climate projections beyond the 21\textsuperscript{st} century are becoming more widely used, the increasing importance of Greenland meltwater for AMOC weakening after 2100 emphasizes the need to incorporate realistic meltwater estimates into these projections. This would also enable probing the robustness of the role of the Arctic Ocean, which played a key role in driving long-term AMOC changes under strong global warming, as well as shaping the AMOC response to Greenland meltwater, in EC-Earth3. This role of the Arctic Ocean is consistent with previous work demonstrating a northward shift of AMOC source regions \cite{Lique2018} and areas of deep convection \cite{Brodeau2016,Lique2018a} as the winter sea ice edge retreats northward under global warming, although only a relatively small number of CMIP6 models (including variants of EC-Earth) shows sustained deep convection in the Arctic Ocean at the end of the 21\textsuperscript{st} century \cite{Heuze2024}. According to our source region analysis, models robustly project a significant Arctic contribution to the AMOC under strong CO$_2$ forcing, but it does not persist until 2300 in most models. This means that EC-Earth3 probably provides an upper bound on role of the Arctic Ocean to meltwater-induced future AMOC weakening.

The absence of abruptness and irreversibility even under strong melt rates of more than \SI{0.3}{Sv} may seem at odds with the robustness of a freshwater-induced AMOC collapse across the model hierarchy \cite{Dijkstra2024a,Rahmstorf2005,VanWesten2023}. There are several possible explanations. First, the results could be specific to EC-Earth3, for example due to model biases. However, the realistic AMOC strength and negative $F_{ovS}$ as in observations increase our confidence that EC-Earth3 is not biased too stable. We also demonstrated that, as in EC-Earth3, the AMOC is reversible under idealized CO$_2$ reversal in most CMIP6 models. Second, even if the AMOC is in a bistable regime under pre-industrial conditions, it is possible that the AMOC shifts outside the bistable regime at very high CO$_2$ concentrations such as those prescribed by the high-end SSP5-8.5 scenario. This possibility calls for a better (conceptual) understanding of the interplay of CO$_2$ and freshwater forcing on the AMOC \cite{Weijer2019,Willeit2024a}, especially under transient forcing.

Finally, it has previously been shown that the same amount of North Atlantic freshwater input has a reduced effect on the AMOC under global warming due to changes in ocean stratification and in the North Atlantic gyre structure  \cite{Swingedouw2015}. We will explore this aspect in more detail in a follow-up study. A corollary is that strong meltwater forcing under a moderate CO$_2$ increase might provide a higher risk for AMOC tipping. However, Greenland melt rates also generally scale with atmospheric temperatures and therefore CO$_2$ concentrations \cite{Lenaerts2015}. This creates a tug-of-war between an increasing amount of meltwater discharge and the weakening ocean sensitivity to meltwater input under global warming, although this relationship could break down on (multi-)millennial time scales if the Greenland ice sheet itself crosses a tipping point \cite{Bochow2023,Sinet2025}.

In addition to comparing the response in different climate models, future studies should assess whether the overturning changes shown here are robust in high-resolution ocean models which resolve mesoscale eddies (e.g., ref. \cite{Li2023a}). Nevertheless, ref. \cite{Martin2023} demonstrated that the magnitude of AMOC weakening due to Greenland meltwater does not depend strongly on the ocean resolution despite different meltwater propagation pathways, increasing our confidence in estimates of meltwater-induced AMOC weakening from CMIP6-class models. Finally, since the uncertainties regarding the impact of Antarctic meltwater on the AMOC are considerable \cite{Li2023a,Shin2024a,Wunderling2024}, simulations assessing the concurrent impacts of Greenland and Antarctic meltwater on the AMOC in long-term scenarios are needed in the future.

\section{Methods}\label{sec11}

\subsection{Model and AMOC characteristics}
We perform model experiments with a state-of-the-art coupled climate model, EC-Earth3 \cite{Doscher2022}, which participated in CMIP6 \cite{Eyring2016}. EC-Earth3 consists of the atmospheric model IFS cy36r4, the land-surface scheme H-TESSEL \cite{Balsamo2009}, the ocean model NEMO 3.6 \cite{Madec2016}, the sea ice model LIM3 \cite{Rousset2015}, and the OASIS3-MCT coupler \cite{Craig2017}. EC-Earth3 is run at its standard resolution for CMIP6, i.e., a horizontal resolution of about \SI{80}{km} (TL255) for the atmosphere and \ang{1} (about \SI{100}{km}) for the ocean, with a grid refinement to 1/\ang{3} in the tropical ocean. In the vertical, 91 levels are used for the atmosphere and 75 levels for the ocean, where layer depths range from \SI{1}{m} near the surface to \SI{200}{m} in the deep ocean.

EC-Earth3 successfully reproduces the most important basin-scale AMOC metrics at the beginning of the scenario runs (2016--2022). At \ang{26}N, the AMOC strength is \SI{17.1(9)}{Sv}, in agreement with the observational value of \SI{16.9(12)}{Sv} from the RAPID array \cite{Johns2023}. In the subpolar North Atlantic, the simulated overturning across the OSNAP-East line is \SI{15.2(8)}{Sv}, again in good agreement with observations (\SI{16.3(6)}{Sv} \cite{Fu2023}). While EC-Earth3 does not capture the overturning across the OSNAP-West line, this contribution is small (around \SI{3}{Sv}) in observations, such that the total subpolar overturning of \SI{12.5(9)}{Sv} in EC-Earth3 is only slightly weaker than observed (\SI{16.7(6)}{Sv} \cite{Fu2023}). An important metric for AMOC stability, the freshwater transport due to overturning at \ang{34}S ($F_{ovS}$), is \SI{-0.014(10)}{Sv} in EC-Earth3. This is slightly larger than in observations (\SI{-0.16(9)}{Sv} \cite{Arumi-Planas2024}) but has the correct sign (negative as observed) often associated with the AMOC being in a bistable regime \cite{DeVries2005,Huisman2010,Weijer2019}. Hence, EC-Earth3 is one the very few CMIP6 models which both simulate a realistic AMOC strength and the correct sign of $F_{ovS}$ \cite{VanWesten2024b}.

\subsection{Experiment setup}
We conduct two ensembles (``reference'' and ``meltwater'') of future projections under the SSP5-8.5 scenario \cite{ONeill2016}, the strongest global warming scenario considered by the IPCC Sixth Assessment Report with unabated greenhouse gas emissions beyond the 21\textsuperscript{st} century \cite{Meinshausen2020}. This scenario can be regarded as a ``worst-case, no-policy'' scenario suitable to study extreme climate outcomes \cite{Hausfather2020}. In SSP5-8.5, CO$_2$ concentrations reach more than \SI{1000}{ppm} (about 4x pre-industrial) by the end of the 21\textsuperscript{st} century and more than \SI{2000}{ppm} (about 8x pre-industrial) by 2200, stabilizing at similarly high levels afterwards \cite{Meinshausen2020}. In EC-Earth3, this leads to a GMST increase of \SI{5.5}{K}, \SI{10.1}{K} and \SI{10.9}{K} compared to pre-industrial at the end of the 21\textsuperscript{st}, 22\textsuperscript{nd} and 23\textsuperscript{rd} centuries, respectively.

The ``reference'' experiments use the standard CMIP6 version of EC-Earth3, except for fixing the surface albedo at the location of present-day ice sheets to 0.8 (the default ice sheet albedo in EC-Earth). This change was implemented because IFS ``parametrizes'' ice sheets as a 10-meter snow pack and exposes low-albedo bedrock after it melts, inducing an unrealistically strong temperature feedback. In comparison, a fixed ice sheet albedo is more in line with the moderate albedo changes even under prolonged strong CO$_2$ forcing \cite{Muntjewerf2021} and enables studying the effect of Greenland meltwater alone, leaving the opportunity for future model experiments to study the effect of changes in the ice cover separately. The ``meltwater'' experiments use the same EC-Earth3 configuration as the ``reference'' runs except for prescribing runoff and calving from the Greenland ice sheet, which is described in the following section.

As is common practice in climate projections \cite{Deser2020}, we use an ensemble of simulations initialized from different initial conditions to separate forced changes from internal variability. The initial conditions are sourced from a subset of the 50-member SMHI Large Ensemble with EC-Earth3 \cite{Wyser2021}. Because low-frequency variability -- and therefore ensemble spread -- in EC-Earth3 is dominated by the AMOC \cite{Doscher2022}, we select four ensemble members that span the range of simulated present-day AMOC strength and projected 21\textsuperscript{st}-century AMOC weakening under SSP5-8.5 (Fig. \ref{fig:ensemble-initialization}a). This includes the two members with the strongest and least pronounced weakening as well as two members with average AMOC characteristics, yielding an overall representative sample (Fig. \ref{fig:ensemble-initialization}b). Both our 4-member ensembles (``reference'' and ``meltwater'') are initialized from the same set of initial conditions at the start of 2016. Note that, due to the slight modification to the model, our reference simulations sample a different realization of internal variability compared to the SMHI Large Ensemble members, so that the rate of weakening is not expected to be identical. However, the strong correlation between the initial AMOC state and 21\textsuperscript{st} century AMOC weakening (Fig. \ref{fig:ensemble-initialization}a) means that we still expect to sample a wide range of rates of AMOC weakening.

\subsection{Meltwater forcing}
In the ``Meltwater'' ensemble, we prescribe runoff and calving from a fully coupled climate--ice sheet model, the Community Earth System Model version 2 (CESM2) coupled to the Community Ice Sheet Model version 2 (CISM2) for the Greenland ice sheet  \cite{Muntjewerf2021}. CESM2 \cite{Danabasoglu2020a} has a nominal resolution of \ang{1} for atmosphere and ocean and CISM2 \cite{Lipscomb2019} was run at \SI{4}{km} resolution. Ice-sheet runoff is routed following topography gradients to the nearest ocean gridpoint, while calving is spread diffusively over a radius of up to \SI{300}{km} from the coast to mimic icebergs \cite{Muntjewerf2020}. Climate projections under the SSP5-8.5 scenario until 2100 were presented in ref. \cite{Muntjewerf2020}.

We prescribe CESM fields for ocean surface fluxes from runoff (monthly) and calving (annually) from 2016 until 2300, including the diffusive spreading of calving. We set grid cells with negative river runoff values, which are artifacts from water conservation in CESM's land model, to zero, such that the total runoff into the ocean matches the runoff from the ice-sheet model well. 
These fluxes were remapped conservatively onto the native NEMO grid before being passed to EC-Earth3 following the implementation of ref. \cite{Devilliers2023}. At coupling time, we set internally calculated runoff and calving over Greenland to zero to avoid double-counting. As in the standard EC-Earth3 configuration, runoff is inserted into the ocean at sea surface temperature, zero salinity, and through a depth of up to \SI{150}{m} to prevent physical and numerical issues from injecting large amounts of runoff into the topmost ocean cell \cite{Madec2016}. In the meltwater ensemble, the runoff depth mask is updated around Greenland to match the runoff climatology by the end of the 23\textsuperscript{rd} century. For calving, we account for the latent heat flux of melting.

The area-integrated time series of runoff over Greenland is shown in Fig. \ref{fig:fw-forcing}a. During the historical period, both the fully coupled ice sheet model in CESM2--CISM2 (19.4 mSv, all values averaged 1981--2010) and the simple mass balance approach in EC-Earth (8.6 mSv) produce a realistic magnitude of runoff compared to observations (13.2 mSv; \cite{Bamber2018}). In the future projections, GrIS runoff strongly increases in CESM2--CISM2, reaching \SI{0.09}{Sv} by the end of the 21\textsuperscript{st} century and more than \SI{0.3}{Sv} by the end of the 23\textsuperscript{rd} century (Fig. \ref{fig:fw-forcing}a). In the EC-Earth reference simulations, Greenland runoff levels off at about \SI{0.07}{Sv} after 2150, a similar value as in many other CMIP6 models (not shown). The meltwater forcing prescribed in the model intercomparison of Bakker et al. \cite{Bakker2016} was also about \SI{0.07}{Sv} in 2300 \cite{Schmittner2016}, such that no significant effect on the AMOC would be expected with their meltwater parametrization. Calving in CISM2 decreases during the 21st century and beyond (\cite{Muntjewerf2020}, Fig. \ref{fig:fw-forcing}a), but the uncertainty of this projection can be considered large because CISM2 does not account for ocean forcing at marine-terminating Greenland margins (cf. \cite{Slater2019}). In any case, the contribution of calving in SSP5-8.5 is expected to be small compared to the strongly increasing runoff.

We note that there are several limitations to our approach. First, we prescribe monthly fields from a single model, such that the meltwater trajectory inherits the boundary conditions from the CESM2 simulation. A recent study comparing long-term projections from a standalone GrIS model forced with boundary conditions from different GCMs showed that CESM forcing produced the largest GrIS mass loss \cite{Greve2022}, implying that the meltwater forcing derived from CESM2-CISM2 is likely at the upper end of the CMIP6 range. Second, our implementation does not account for the differences in surface mass balance between CESM2 and EC-Earth3, such that the sea level rise in the EC-Earth3 ``meltwater'' runs is different from the integrated CISM2 mass loss. However, we do not analyze sea level metrics here, making this an acceptable trade-off over ad-hoc corrections. We preferred the forcing approach from a fully coupled model over parametrizations \cite[e.g.,][]{Fettweis2013,Lenaerts2015,Bakker2016} for several reasons: the extended SSP5-8.5 forcing extends far beyond their calibration range; parametrizations typically give regional averages while CESM2 runoff is resolved at \ang{1} resolution; they do not eliminate the role of model biases, as the (regional or global) climate model forcing used for calibration can substantially affect the runoff sensitivity to warming \cite{Little2016}; and they also do not take into account differences in surface mass balance. As a final caveat, fixing the ice sheet geometry and albedo to present-day values allows to cleanly separate the effect of meltwater, but these changes are expected to further weaken the AMOC \cite[e.g.,][]{Davini2015,Andernach2025}. However, the effect is expected to be of second order compared to the effects of CO$_2$ and meltwater.

\subsection{Overturning and source region diagnostics}
First, we analyze the AMOC in depth-space at the conventional latitude of \ang{26.5}N by taking the depth-maximum of the annual mean overturning streamfunction \cite{Buckley2016}. At subpolar latitudes, density ($\sigma$) coordinates are preferred over depth ($z$) coordinates because isopycnals tend to be strongly sloped in east-west direction \cite{Lozier2019}. Following ref. \cite{Zhang2010a}, we therefore compute the overturning north of \ang{40}N in density coordinates referenced to the surface ($\sigma_0$). The streamfunction is then defined as \cite{Buckley2023}
\begin{equation}
    \Psi(\sigma) = - \int_{x_1}^{x_2} \int_{\sigma_\text{bot}}^{\sigma} v(x,\sigma^{\prime},t) \dd \sigma^{\prime} \dd x,
    \label{eq:overturning-density-space}
\end{equation}
where $v$ is the northward velocity, $x_1$ and $x_2$ are the western and eastern boundaries of the basin, and $\sigma_\text{bot}$ is the bottom (maximum) density. Using this definition, the overturning is zero at the bottom and equal to the net volume transport for $\sigma \to 0$. Following previous modeling studies \cite{Menary2020a,Jackson2023b}, we do not apply any compensation term between neighboring straits (except for the comparison against observed values above). We compute $\Psi(\sigma)$ at \ang{45}N, two sections east and west of Greenland that approximately follow the OSNAP array \cite{Lozier2019}, the Greenland-Scotland Ridge (GSR), Davis Strait, Fram Strait and the Barents Sea Opening (Fig. \ref{fig:convergence-sigma-moc}d). Streamfunctions are computed from monthly data using an adapted version of the ``line method'' from the StraitFlux package \cite{Winkelbauer2024a}.

Following ref. \cite{Jackson2023b}, we define the convergence of overturning $M$ in a region as the difference between $\Psi(\sigma)$ at its southern and northern boundaries. This way, the overturning across \ang{45}N is decomposed into the sum of convergences of the five regions north of \ang{45}N (Fig. \ref{fig:convergence-sigma-moc}d; SPG: subpolar gyre, IS: Irminger Sea, LS: Labrador Sea, GIN: Nordic Seas, AO: Arctic Ocean and Baffin Bay):
\begin{equation}
    \Psi_{45N}(\sigma) = M_{SPG}(\sigma) + M_{IS}(\sigma) + M_{LS}(\sigma) + M_{GIN}(\sigma) + M_{AO}(\sigma).
    \label{eq:convergence-45n}
\end{equation}
Here, we assume that the overturning across Bering Strait is zero and the contributions from other marginal seas (Baltic Sea, Hudson Bay etc.) are negligible, which is confirmed by the very good agreement between the left-hand and right-hand side of Eq. \ref{eq:convergence-45n} (not shown).

Similarly to ref. \cite{Buckley2023}, in our analysis we focus on the (time-dependent) density $\sigma_{\text{MOC}}$, which is defined as the density at which the overturning at \ang{45}N is at its maximum. This can be interpreted as the density which bounds the AMOC lower limb. Evaluating Eq. \ref{eq:convergence-45n} at $\sigma_{\text{MOC}}$ therefore quantifies the net contribution of each region to the AMOC lower limb, regardless of which processes drive diapycnal transformations.

\subsection{Volume and buoyancy budget}
If a region is characterized by inflow of lighter and outflow of denser water masses, the overturning can be related to the transformation of water masses at the surface in the so-called Walin framework \cite{Walin1982}. The volume and buoyancy budget in a region can be computed as \cite{Buckley2023}
\begin{align}
    F(\sigma) + G(\sigma) = \frac{\partial V(\sigma)}{\partial t} + \Psi_S(\sigma) - \Psi_N(\sigma),
    \label{eq:buoyancy-budget}
\end{align}
where $F(\sigma)$ is the surface-forced water mass transformation defined below, $V(\sigma)$ is the ocean volume with a potential density larger than $\sigma$, $\Psi_S(\sigma)$ and $\Psi_S(\sigma)$ are the overturning across the southern and northern gateways of the region, respectively, calculated using Eq. \ref{eq:overturning-density-space}. For the Arctic Ocean, $\Psi_N(\sigma)=0$ because there is no overturning across Bering Strait. $G(\sigma)$ quantifies interior mixing which cannot be diagnosed directly from climate model output and is therefore calculated as a residual \cite{Buckley2023}.

Surface-forced water mass transformation $F(\sigma)$ is given by 
\begin{align}
    F(\sigma) = \frac{1}{\Delta \sigma} \iint_{\mathcal{A}} \Bigg[ -\frac{\alpha}{C_p}\,Q + \beta \frac{S}{1-S}\Phi_{FW} \Bigg] \phantom{\Bigg|}\Pi(x,y; \sigma)\phantom{\Bigg|} \dd x \, \dd y,
    \label{eq:sfwmt}
\end{align}
where $\alpha$ and $\beta$ are the thermal expansion and haline contraction coefficients, respectively, $C_p$ is the specific heat capacity of seawater, $Q$ is the air-sea heat flux, $S$ is the sea surface salinity, $\Phi_{FW}$ is the surface freshwater flux and $\Pi(x,y; \sigma)$ selects the outcrop region of a density range $\sigma\pm\frac{\Delta\sigma}{2}$. It is defined as
\begin{align}
    \Pi(x,y; \sigma) = \begin{cases}
        1 & \text{if  } |\tilde{\sigma}(x,y)-\sigma| \leq \frac{\Delta\sigma}{2} \\
        0 & \text{elsewhere},
    \end{cases}
    \label{eq:outcrop-region}
\end{align}
where $\tilde{\sigma}(x,y)$ is the potential density at the location $(x,y)$. $F$ is analogous to a streamfunction and computed from monthly fields for surface density, heat and freshwater fluxes (evaporation minus precipitation minus runoff minus sea-ice melt) and subsequently averaged into an annual mean climatology. A spacing of $\Delta \sigma = \SI{0.1}{\kilogram\per\cubic\meter}$ is used for all budget terms in Eq. \ref{eq:buoyancy-budget}.

\subsection{Time of emergence}
For any given year, the difference in ensemble means between the reference and meltwater ensembles is compared using a one-sided, paired Student's $t$-test. % (\texttt{ttest\_rel} from the Scipy library).
Here, two simulations initialized from the same initial condition but subject to different meltwater trajectories are treated as paired observations due to the long memory in the AMOC time series arising from low-frequency (centennial) variability \cite{Meccia2023,Mehling2024a}. Since the AMOC weakening in all ensemble members is approximately linear in until at least 2120, the $t$-test is first applied to the linear AMOC trend (2016 to the specified year), and a sensitivity test is performed using 15-year running means instead of the trend. The year of emergence is then defined as the first year from which on the difference is always statically significant ($p<0.05$).

For an uncertainty estimate of emergence times, we repeat this procedure on an ensemble of 1000 surrogate time series for each ensemble member. To construct the surrogates, the interannual-to-decadal variability (``residuals'') is obtained by subtracting a fifth-order polynomial fit from the original AMOC time series. Then, the residuals are resampled using the ``random phasing'' method of Ebisuzaki \cite{Ebisuzaki1997}, which preserves the weak autocorrelation of the residuals better than conventional bootstrapping, and added back to the polynomial fit.

\subsection{CMIP6 model data}
To put the results from EC-Earth3 in context, we use output from the CMIP6 multi-model ensemble \cite{Eyring2016} for two experiments: extended SSP5-8.5 scenario simulations until 2300 from ScenarioMIP \cite{Meinshausen2020} and reversibility experiments (\textit{1pctCO2} and \textit{1pctCO2-cdr}) from the Carbon Dioxide Removal Model Intercomparison Project (CDRMIP) \cite{Keller2018}.

The SSP5-8.5 scenario runs use the same external forcing as the EC-Earth simulations (except for not prescribing meltwater) and we select all 8 models (cf. Fig. \ref{fig:convergence-arctic-cmip6}) which provide sufficient ocean output until 2300 to calculate the overturning in density space.

The CDRMIP experiments use an idealized CO$_2$ ramp-up/ramp-down protocol starting at pre-industrial conditions. First, CO$_2$ concentrations are increased by 1\% per year (``ramp-up'') until they reach four times pre-industrial values (around \SI{1140}{ppm}) after 140 years. Subsequently, CO$_2$ concentrations are decreased at 1\% per year (``ramp-down'') for another 140 years until they return to pre-industrial values. The simulations are then extended for a minimum of 50 years under fixed pre-industrial conditions (``stabilization''). A total of 8 CMIP6 models listed in Fig. \ref{fig:reversibility} provided AMOC output for this experiment.

\backmatter

\clearpage
\nolinenumbers

\section*{Supplementary Figures}

\renewcommand\thefigure{S\arabic{figure}}
\setcounter{figure}{0}

\begin{figure}[htbp]
    \centering
    \includegraphics[width=\textwidth]{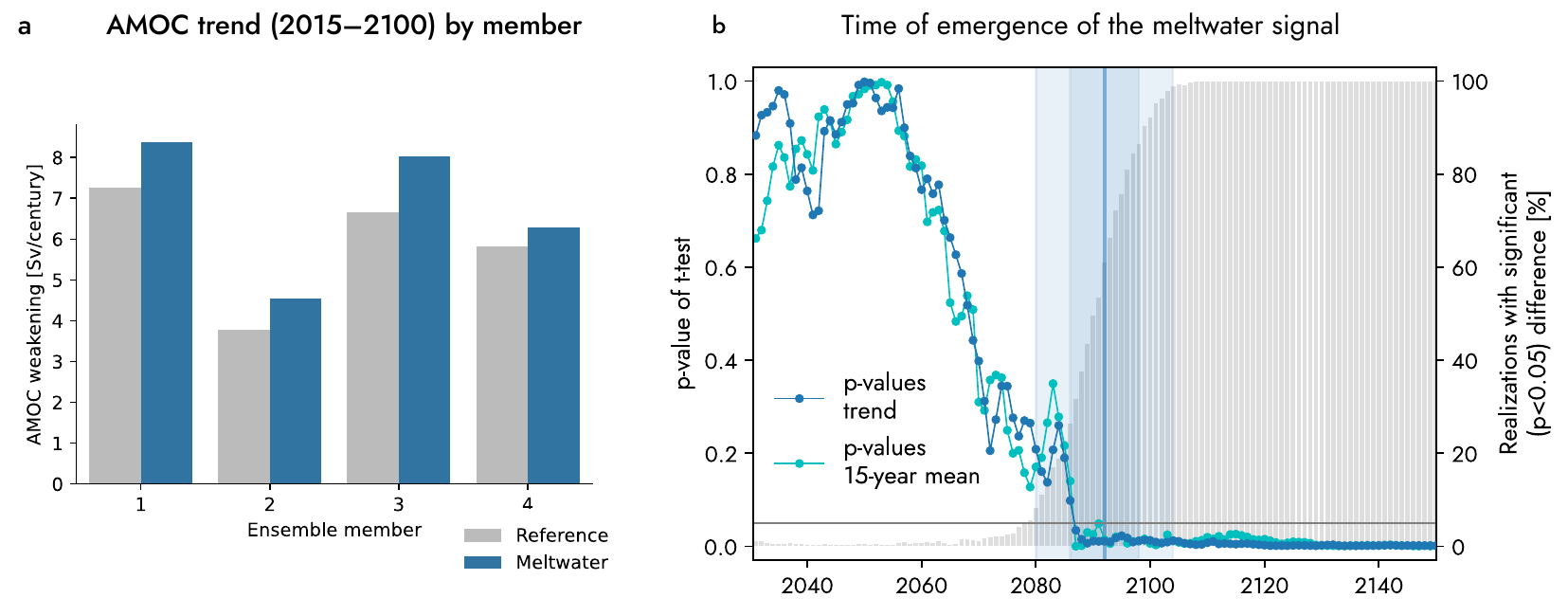}
    \caption{\textbf{21st-century AMOC weakening and emergence of the meltwater signal.} (a) Linear trends of AMOC weakening (in Sv/century) during 2015--2100 for each ensemble member. Paired bars are initialized from the same initial conditions. (b) Left axis: $p$-values for the one-sided $t$-test whether trends (blue line) or 15-year running means (cyan line) differ significantly between the reference and meltwater ensembles. The grey line shows the 0.05 threshold. Right axis: Percentage of realizations using a bootstrap test (Methods) that show a significant difference in trends between the reference and meltwater ensembles. The blue line and shadings indicate the median, 66\% and 95\% confidence intervals of the time of emergence (Methods).}
    \label{fig:amoc-trends-21st}
    \hypertarget{fig:s1}{}
\end{figure}

\begin{figure}[htbp]
    \centering
    \includegraphics[width=\textwidth]{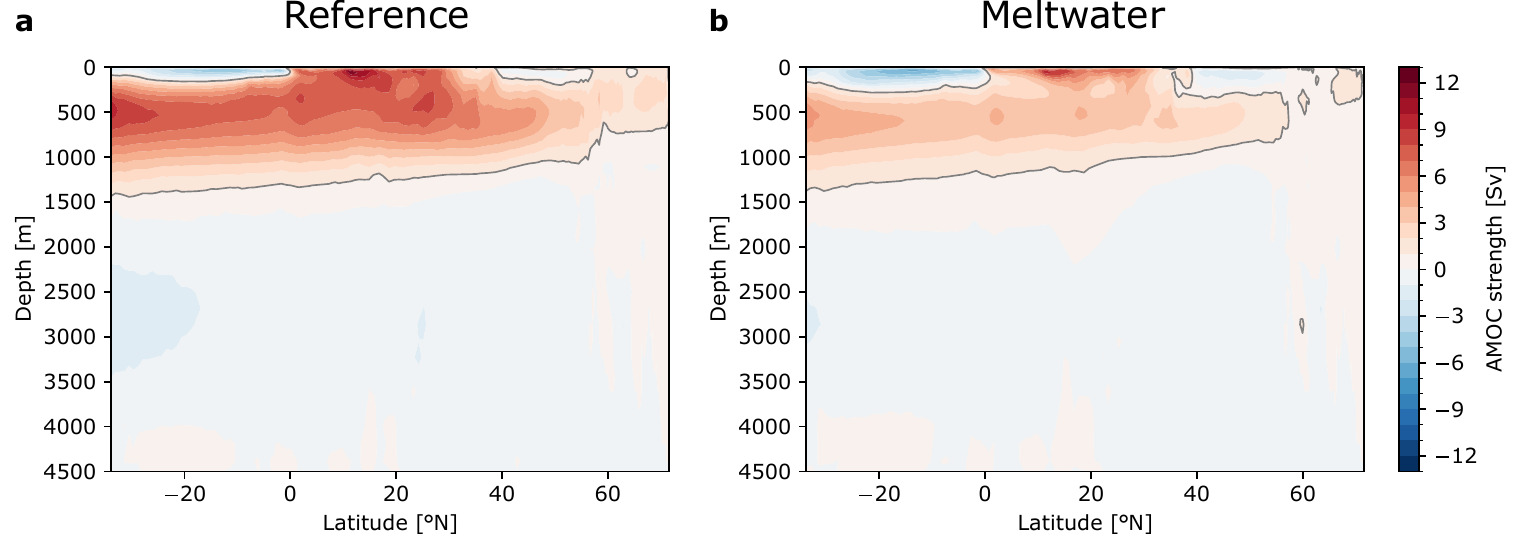}
    \caption{\textbf{Final AMOC streamfunctions.} AMOC streamfunctions in depth space for the end of the simulation 2270--2300: (a) Reference (2 members), (b) Meltwater (1 member). The grey contour indicates the \SI{1}{Sv} isoline.}
    \label{fig:amoc-streamfunctions-23rd}
\end{figure}

\begin{figure}[htbp]
    \centering
    \includegraphics[width=0.7\textwidth]{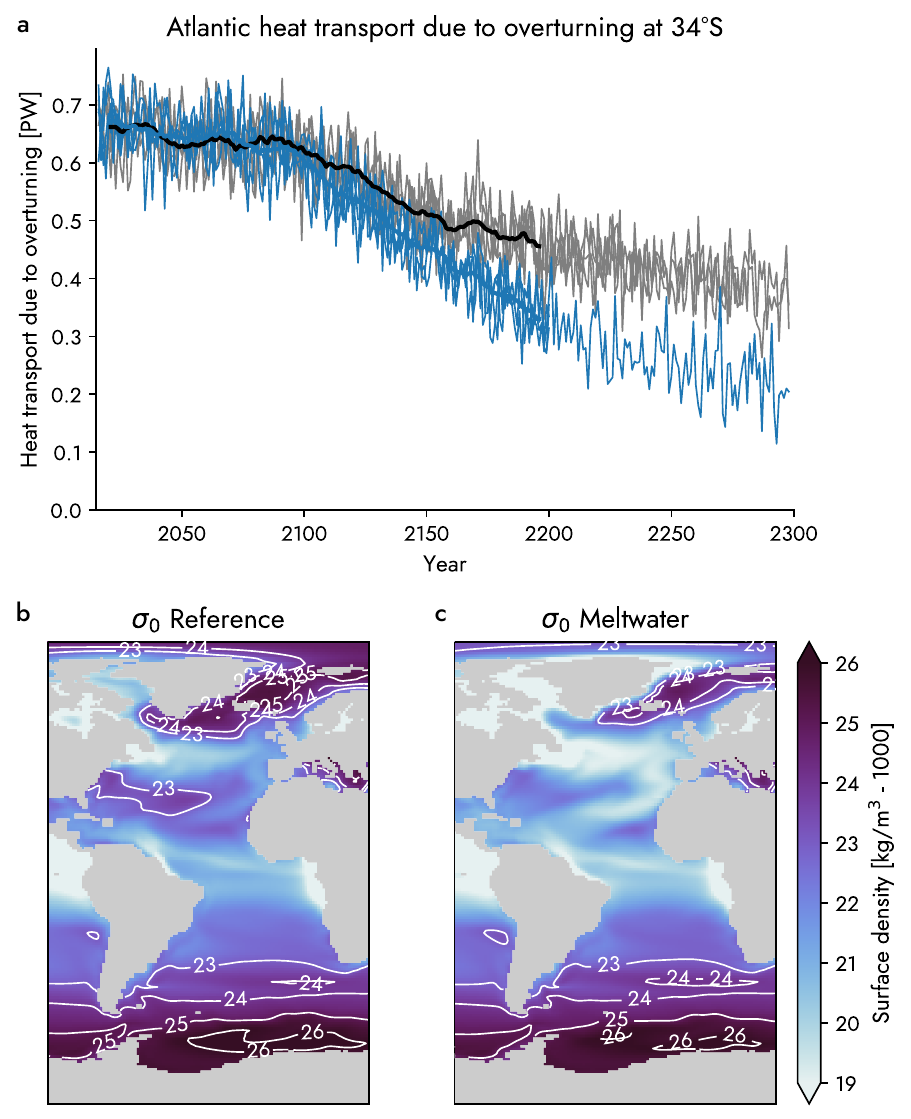}
    \caption{\textbf{Characteristics of a sustained AMOC.} (a) Time series of Atlantic heat transport due to overturning (Methods) for the reference and meltwater ensembles, (b, c) Annual mean surface density ($\sigma_0$) at the end of the 23\textsuperscript{rd} century for the reference and meltwater simulations. Shared isopycnals between the North Atlantic and Southern Ocean are shown as white contours.}
    \label{fig:characteristics-amoc-2300}
\end{figure}

\begin{figure}[htbp]
    \centering
    \includegraphics[width=\textwidth]{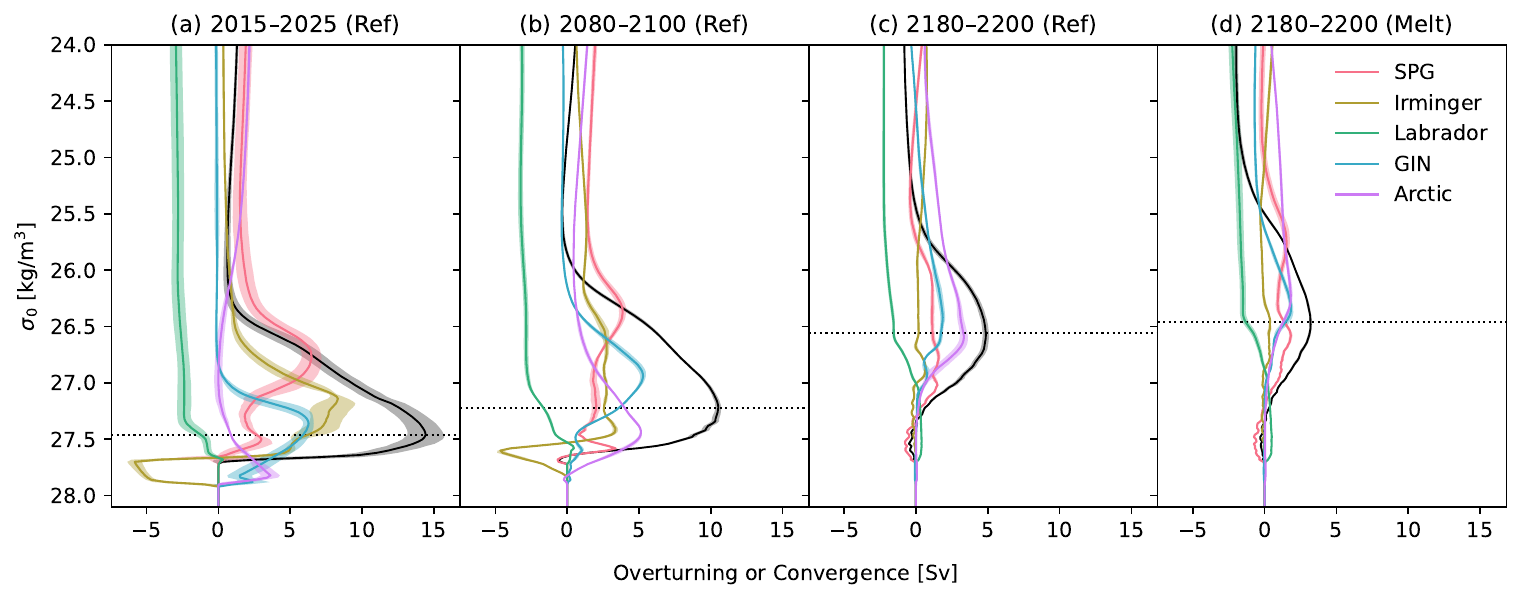}
    \caption{\textbf{Changing overturning convergence by region.} Climatologies of $\sigma$-space overturning at \ang{45}N (black lines) and convergence by region. Shadings indicate the ensemble standard deviation. The dotted horizontal line indicates $\sigma_\text{MOC}$, the density at which the overturning streamfunction at \ang{45}N is at its maximum.}
    \label{fig:moc-convergence-snapshots}
\end{figure}

\begin{figure}[htbp]
    \centering
    \includegraphics[width=\textwidth]{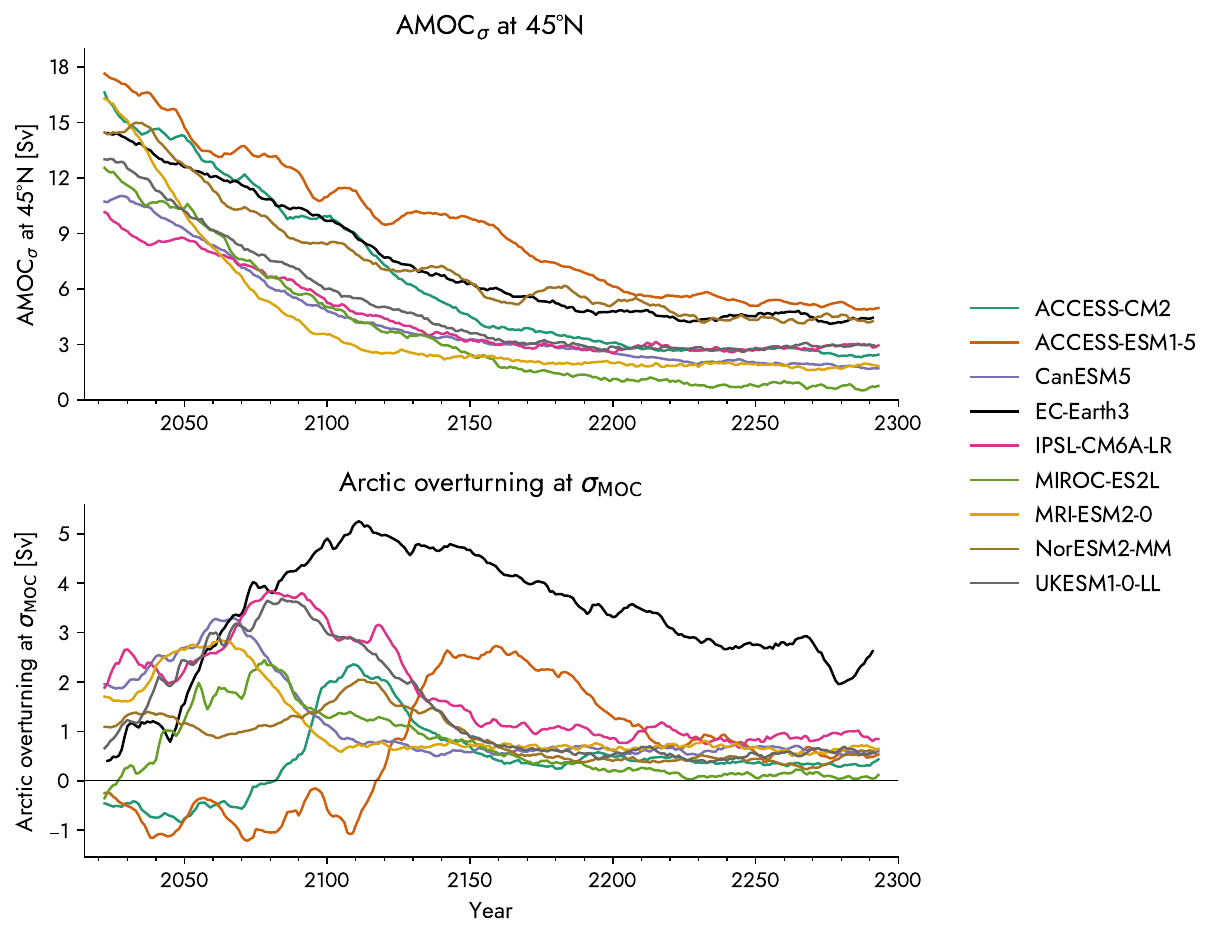}
    \caption{\textbf{Mid-latitude and Arctic overturning in CMIP6 models.} (a) Maximum Atlantic overturning in density-space at \ang{45}N, (b) overturning across the Arctic gateways at $\sigma_\text{MOC}$ in the SSP5-8.5 scenario for all CMIP6 models providing suitable output until 2300 and the first ensemble member of the EC-Earth3 ``Reference'' ensemble. A 15-year running mean is applied to all time series for clarity.}
    \label{fig:convergence-arctic-cmip6}
\end{figure}

\begin{figure}[htbp]
    \centering
    \includegraphics[width=\textwidth]{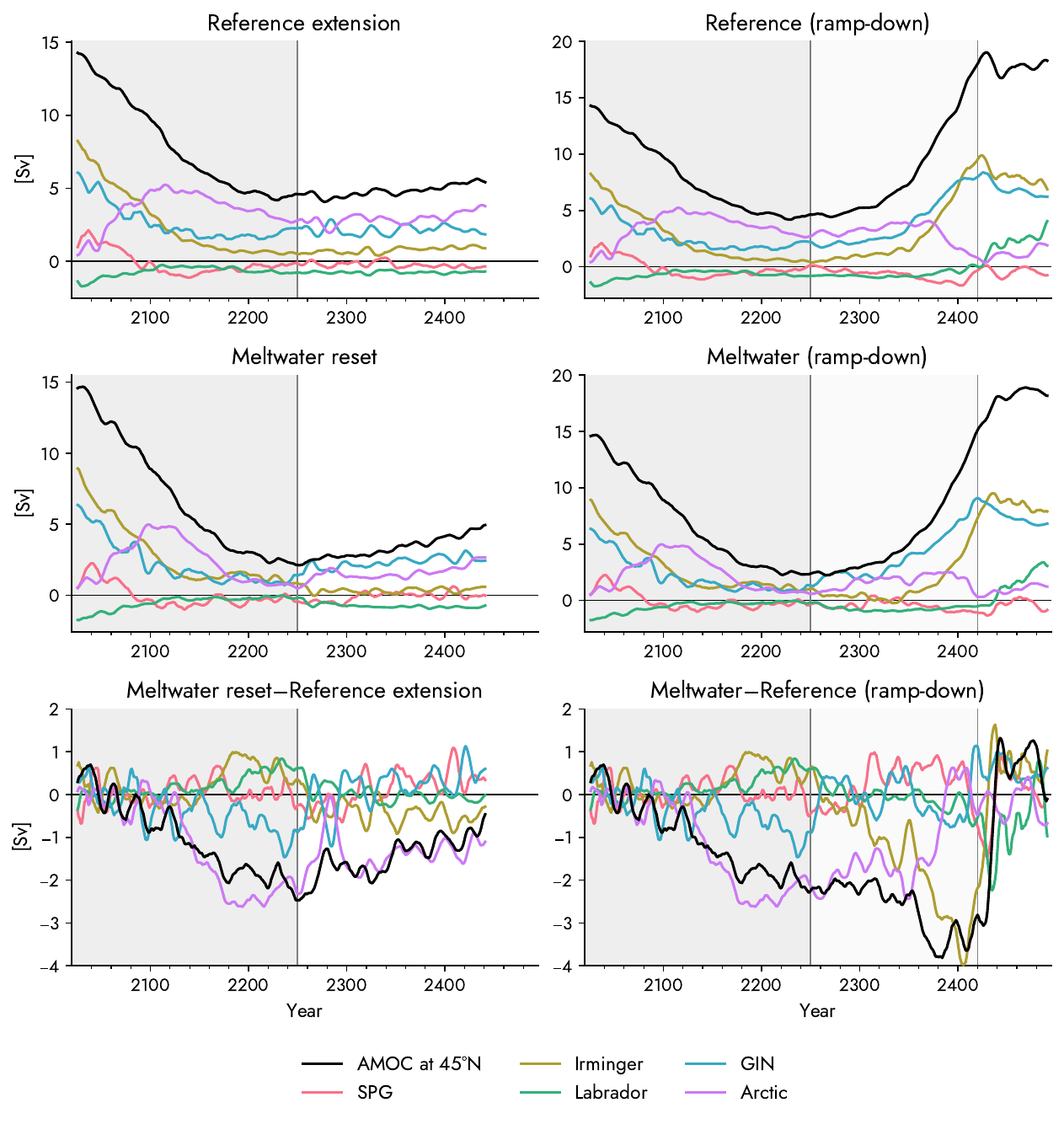}
    \caption{\textbf{Convergence at $\sigma_\text{MOC}$ in EC-Earth3 reversibility experiments.} Same as Fig. \ref{fig:convergence-sigma-moc}, but for the reversibility experiments shown in Fig. \ref{fig:reversibility}.}
    \label{fig:convergence-reversibility}
\end{figure}

\begin{figure}[htbp]
    \centering
    \includegraphics[width=0.49\textwidth]{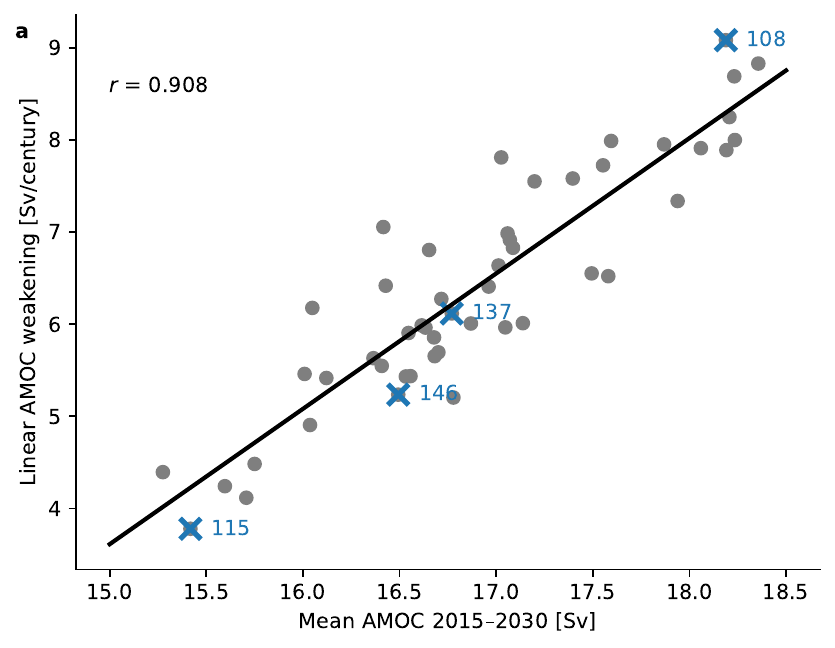}\hfill
    \includegraphics[width=0.49\textwidth]{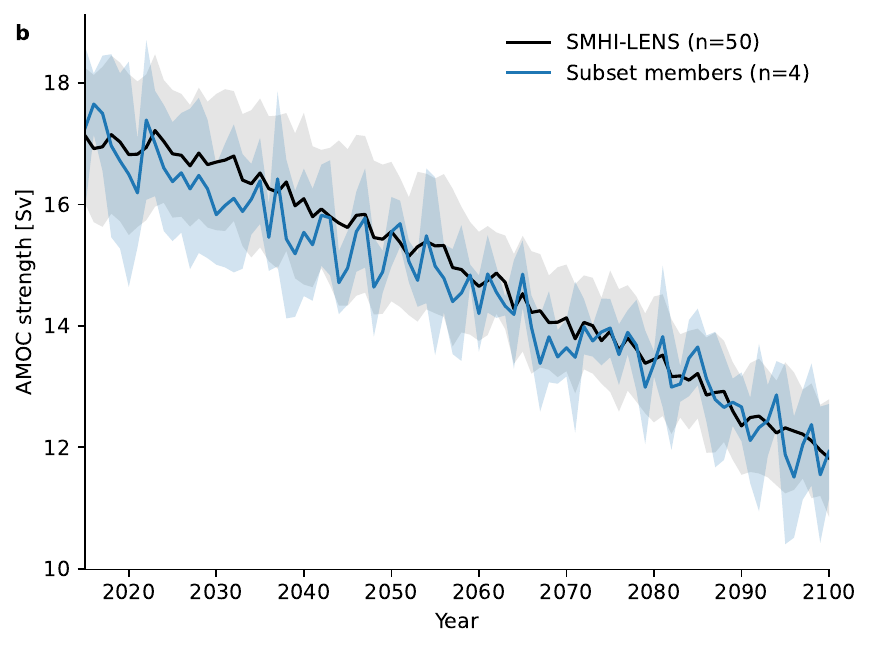}
    \caption{\textbf{Ensemble member selection strategy} based on the SMHI Large Ensemble. (a) Relation between initial (2015--2030) mean AMOC strength at \ang{26}N and linear AMOC trend during 2015--2100 in the 50-member EC-Earth3 large ensemble \cite{Wyser2021} for the SSP5-8.5 scenario. The selected ensemble members and their original variant labels (rXXXi1p1f1) are highlighted with blue crosses. (b) AMOC time series (ensemble mean plus and minus one standard deviation) under SSP5-8.5 forcing for the full 50-member ensemble (black) and the 4-member subset (blue). While the subset mean is slightly lower than the ensemble mean between the 2020s and 2040s, overall the subset captures the ensemble mean and spread well, especially in the second half of the 21\textsuperscript{st} century.}
    \label{fig:ensemble-initialization}
\end{figure}

\FloatBarrier

\linespread{1.0}\selectfont

\end{document}